# Threshold behavior of free electron lasers without inversion

Dmitry N. Klochkov [1], Ashot H Gevorgyan [2], Koryun B. Oganesyan [3*],

Nerses S. Ananikian [3], Nikolay Sh. Izmailian [3], Yuri V. Rostovtsev [4], Gershon Kurizki [5]

[1] General Physics Institute RAS, Moscow, Russia

[2] Yerevan State University, Yerevan, Armenia

[3] Yerevan Physics Institute, Alikhanyan National Science Lab,

Alikhanyan Br. 2, Yerevan 0036, Armenia

[4] Department of Physics, University of North Texas, Denton, TX 76203, USA

[5] Chemical Physics Department, Weizmann Institute of Science, Rehovot 76100, Israel

[*] bsk@yerphi.am

The interaction between noncolinear laser and relativistic electron beams in static magnetic undulator has been studied within the framework of dispersion equations. For a free-electron laser without inversion (FELWI), the threshold parameters are found. The large-amplification regime should be used to bring an FELWI above the threshold laser power.

## 1. Introduction

Usually FEL [1,2] use the kinetic energy of relativistic electrons moving through a spatially modulated magnetic field(wiggler) to produce coherent radiation. The frequency of radiation is determined by the energy of electrons, the spatial period of magnetic field and the magnetic field strength of the wiggler. This permits tuning a FEL in a wide range unlike atomic or molecular lasers. However for purposes of achievement of short-wavelength region of generation there are important possible limitations of the FEL gain.

The idea of inversionless FEL or FELWI (FEL without inversion) was formulated and discussed by M.O. Scully and coworkers [3-7]. In the usual FEL the gain $G$ is an antisymmetric function of the detuning $\Delta = E - E_{res}$, where $E$ and $E_{res}$ are the electron energy and its resonance value in the undulator. The integral of such a gain over $\Delta$ (or $E$) is equal to zero. By definition, in the FELWI $\int G(E)dE \neq 0$ and mainly, $G > 0$. Moreover, if in the usual FEL in the "hot-beam" regime (i.e., in the case of a broad electron energy distribution) the averaged gain is proportional to the squared inverse width of the distribution function, $(\delta E)^{-2}$, in FELWI $G(E) \propto (\delta E)^{-1}$. Hence, in the case of energetically wide beams the FELWI gain can exceed significantly the gain of the usual FEL. This advantage of FELWI (compared to usual FELs) makes such devices particularly interesting and potentially perspective in short-wave-length regions.

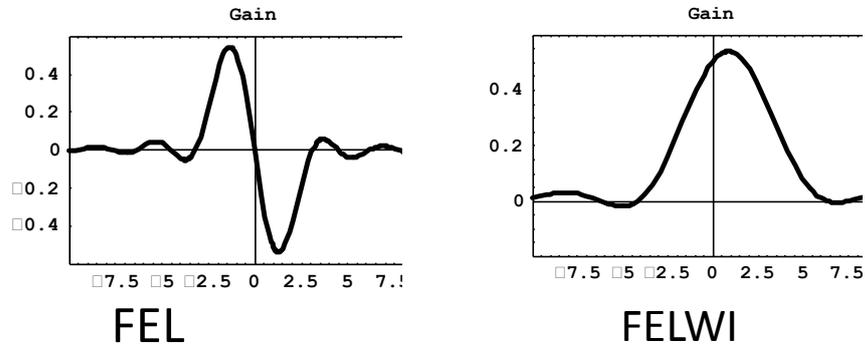

Fig.8. The Gain vs detuning, $\Omega \equiv \omega(v_0 - v_{res})/c$, which characterizes deviation of the electron velocity or the laser frequency from the resonance condition for usual FEL (left) and for FELWI (right).

The conditions $\int G(E)dE \neq 0$ and $G > 0$ imply that amplification of light can take place almost at any position of the resonance energy $E_{res}$ with respect the energy $E = E_0$, at which the electron distribution function $f(E)$ is maximal. In FELWI amplification can take place both at positive and negative slopes of the function $f(E_{res})$, as well as its peak. This feature of FEWI is in a great contrast with that of FEL, where amplification can take place only at the positive slope $f(E_{res}) > 0$. This last condition is easily interpreted as the condition of inversion: in FEL the number of electrons with $E > E_{res}$ must be larger than with $E < E_{res}$. In FELWI amplification can occur independently of the relation between $E_{res}$ and $E_0$. This means that for amplification in FELWI it does not matter whether the number of particles with energy $E > E_{res}$ is larger than with energy $E < E_{res}$ or not. This explains an origin of the concept "without inversion" for the kind of FEL to be considered.

More specifically, the idea of FELWI is based on a two-wiggler scheme with a specially organized dispersion region between the wigglers. In principle, the two-wiggler scheme is widely used in normal FELs. This scheme (often referred to as an optical clystron) is known to provide a somewhat higher gain with narrower amplification band than in a single-wiggler FEL but it does not provide conditions for amplification without inversions. The reason is in features of devices between two wigglers of a FEL. In all the existing two-wiggler FEL these devices (or no special devices at all) are the so-called positive-dispersion devices. This means that the higher-energy electrons of a beam cross a space between the first and second wigglers in a shorter time than the lower-energy ones. For creation of FELWI, a device between the wigglers must be rather unusual: it must provide the negative-dispersion regime in which the faster electrons spend longer time in the dispersion region than the slower ones. This goal does not look unachievable but this is not easy to reach it.

A concept of FELWI is related to that of Lasing Without Inversion (LWI) [8] in atomic systems: three-level systems or systems with autoionizing atomic levels. In both cases effects of amplification without inversion are explained by interference. But specific kinds of interference in FELWI and atomic systems are significantly different. In atomic systems amplification without inversion is attributed to interference of different channels of transitions between the same initial and final atomic states. In contrast, interference in FELWI has a purely classical character. A typical scheme of FELWI involves two wigglers and a dispersion zone between them. Field-induce corrections to classical electron trajectories acquired in the first and second wigglers interfere with each other. A proper construction of the dispersion zone can give rise to such a form of interference which provides the described above spectral features of the FELWI gain (for more details on FEL see [16-82]).

## 2. The Threshold and Deviated Angle

The basic idea of FELWI [3-7,9,10] is that the electrons, that have passed the first undulator have a dispersion of the transverse velocity, and as a consequence of the angle between the vector of velocity and the wiggler axis, and this dispersion is directly connected to final gain of energy. Therefore the selection of a direction of electron motion is equivalent to the selection of energy, that basically allows to change in a controllable way the length of the drift region of electrons with different energies. This mechanism can work only if the spread of the angle α, arising as a result of interaction of electrons with the field in the first wiggler of FELWI, is larger than the natural dispersion of the directions in the electron beam, $\Delta\alpha_{beam}$. This circumstance leads to the occurrence of the threshold for laser radiation power at the point of entry of the FELWI's first wiggler.

The interaction of electron beam with laser field can be described by laws of conservation for momentum $\mathbf{p}_e + \mathbf{p}_L = \mathbf{p}_e' + \mathbf{p}_L'$ and energy $\varepsilon_e + \varepsilon_L = \varepsilon_e' + \varepsilon_L'$. Here $\mathbf{p}_e$ and $\mathbf{p}_e'$ are initial and final momentums of electrons, $\mathbf{p}_L$ and $\mathbf{p}_L'$ are initial and final momentums of laser field; $\varepsilon_L$ and $\varepsilon_{L'}$ are initial and final energies of light beam and $\varepsilon_e$ and $\varepsilon_{e'}$ are initial and final energies of electrons. The density of electromagnetic wave momentum is $P_L = (1/4\pi c)[\mathbf{EB}] = \mathbf{k}\omega/(4\pi c)\mathbf{A}_L^2$, where $\mathbf{A}_L$ is an amplitude of a vector-potential of laser field.

We can write for $\mathbf{A}_{L'} = \mathbf{A}_L \exp(k''L)$, where $k''$ is a spatial growth rate of laser field in a medium of an electron beam; $L$ is a length of interaction. From law of conservation we can expect that $|\Delta p| = |p_e' - p_e| = |p_L' - p_L| = \mathbf{A}_L^2 \exp\left[(2k''L) - 1\right]$. We can see that the change of electron momentum $|\Delta p|$ depends on the spatial growth rate $k''$: with the growth rate $k''$ rising, the change of electron momentum rises too. This means that for noncolinear interaction the deviation of electron from its original direction depends on both the spatial growth rate $k''$ and the initial amplitude $A_L$ of laser field. The growth rate $k''$ is a function on electron beam current; and the amplitude depends on laser power at the entrance of undulator. Therefore, the condition $\alpha > \alpha_{beam}$ leads to the threshold of either the laser power at the entrance of undulator or the electron beam density.

We consider the induced radiation by a mono-energetic beam of electrons propagating in a wiggler. We assume that the static magnetic field of a plane undulator $\mathbf{A}_w$ is independent on the transverse coordinates $x$ and $y$. Also we approximate the static magnetic field by a harmonic function $\mathbf{A}_w = A_w \mathbf{e}_y = \left( A_0 e^{-i\mathbf{k}_w \mathbf{r}} c.c. \right) \mathbf{e}_y$, where $\mathbf{k}_w = (0, 0, k_w)$ is the wiggler wave vector; ``c.c." denotes the complex conjugation, $\mathbf{e}_y$ is the unit vector along $y$ axis. The wiggler field causes an electron to oscillate along the $y$-axis. For this reason, the electron interacts most efficiently with a light wave if the latter is linearly polarized. We assume that the vector potential of the laser wave has a linear polarization $\mathbf{A}_L = A_L(t, x, z) = a^+ e^{i(\mathbf{k}-\mathbf{k}_w)\mathbf{r} - i\omega t}$.

The dispersion equation of EM oscillations in the plasma like electron beam medium is an algebraic equation of power four for wave vector $\mathbf{k}$ (see equation (2) in [11]). Therefore, there are four solutions $k_j$ describing two beam waves, slow and fast, having the forms $k = \omega/u \pm \sigma\omega_b$ ($\sigma$ is a shape factor), and two electromagnetic waves, one of which extends to the direction of the beam moving while another propagates to the opposite direction. The solution of the linearized equations for slow motion of the electron in the $xz$-plane is [11,12]:

$$\delta \mathbf{v}_\square = K^2 \frac{c^2}{\gamma_0^3} \sum_{j=1}^{4} \frac{\beta_{1j}\mathbf{k}_j - \dfrac{\omega}{c^2}\beta_{2j}u}{D_{b(j)}} a_j e^{i\mathbf{k}_j \mathbf{r}_\square - i\omega t} + c.c \qquad (1)$$

Here $a_j = a_{+j}/A_0$ is the dimensionless initial amplitude of wave $j$ with wave vector $k_j$, j numbers four branches of oscillations in the electron beam medium: two laser waves and two beam waves, $\vec{\mathbf{u}} = (-u\sin\alpha; 0; u\cos\alpha)$ is the electron velocity; ``c.c." denotes the complex conjugation. $D_b = (\omega - \mathbf{k}\mathbf{u})^2 - \Omega_b^2$ is the dispersion function of electron beam wave associated with the beam frequency $\Omega_b$, where $\Omega_b^2 = \omega_b^2 \left[ 1 - (\mathbf{k}\mathbf{u})^2/(kc)^2 \right]/\gamma_0$. Here $\omega_b^2 = 4\pi e^2 n_b/m$ is square of the Langmuir frequency of the electron beam corresponding density $n_b$. $K$ is the undulator strength parameter, defined as normalized dimensionless vector-potential of the undulator magnetic field $K = \dfrac{e}{mc^2}|A_0|$. The total relativistic factor of electrons $\gamma_0$ is defined as

$\gamma_0 = \sqrt{1+2K^2}\left(1-u^2/c^2\right)^{-1/2}$. The coefficients $\beta_1$ and $\beta_2$ are

$\beta_1 = \gamma_0\left(\omega-(k_0 u)\right)-\omega_b^2(k_0 u)/(k_0 c)^2$ and $\beta_2 = \gamma_0\left(\omega-(k_0 u)\right)-\omega_b^2/\omega$.

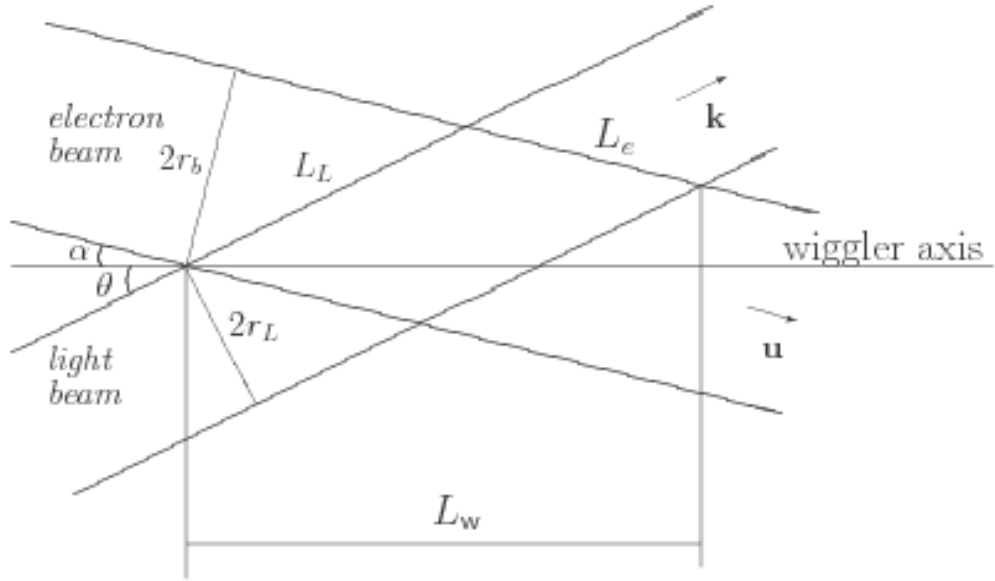

Fig.1. The scheme in $xz$-plane of undulator with non-collinear arrangement.

The equation of a trajectory (1) is obtained for mono-energetic electron beam having the unlimited size and, therefore, indefinitely long interacting with a electromagnetic field. Below we take into account the finite size of electron and laser beams. At first, the non-collinear arrangement of electron and laser beams leads to the finite area of them interaction. The length of laser amplification in the medium of electron beam is $L_L = 2r_b/\sin(\alpha+\theta)$. Here $2r_b$ is a width of the electron beam in the $xz$-plane. The length, at which the electrons move acting by force of laser field, is equal to $L_e = 2r_L/\sin(\alpha+\theta)$. Here $2r_L$ is a width of the laser beam in the $xz$-plane. The working length of the wiggler is $L_w = L_e \cos\alpha + L_L \cos\theta \approx 2(r_L+r_b)/\sin(\alpha+\theta)$.

Second, at the entrance of undulator the perturbation of the velocity is absent: $\delta v = 0$, that leads to

$$\sum_{j=1}^{4} \frac{\beta_{1j}\mathbf{k}_j - \frac{\omega}{c^2}\beta_{2j}\mathbf{u}}{D_{b(j)}} a_j e^{i\mathbf{k}_j \mathbf{r}_0 - i\omega t} = 0 \tag{.2}$$

Using Eq.(2) we can rewrite the Eq.(1) in form

$$\delta \mathbf{v}_\| = K^2 \frac{c^2}{\gamma_0^3} \sum_{j=1}^{4} \frac{\beta_{1j}\mathbf{k}_j - \frac{\omega}{c^2}\beta_{2j}\mathbf{u}}{D_{b(j)}} a_j e^{i\xi_0}\left(e^{-i\Delta_{\omega(j)}t}-1\right) + c.c \qquad (3)$$

Where $\Delta_{\omega(j)} = \omega - (\mathbf{k}_j \mathbf{u})$ is the detuning, $\xi_0 = \mathbf{k}_0 \mathbf{r}_{\Box 0}$, and $\mathbf{r}_{\Box 0}$ is the initial coordinate in the *XZ* plane. $a_j$ are the amplitudes of waves at the point of entry of the FELWI's first wiggler. For the problem of spatial amplification, when the spatial growth rate is considered, the initial amplitudes $a_j$ are free parameters. Other words say, we consider the problem of a laser amplifier for the first section of such a FEL.

Assuming that the electron, incoming in the laser field at $t = 0$ and interacting during the time $t = L_e / u$, deviates from the initial direction by the angle $\Delta\alpha$, we obtain

$$\delta\alpha = \tfrac{\delta v_\perp}{u} = 2K^2 \frac{c^2}{\gamma_0^3} \mathrm{Re}\left\{ \sum_{j=1}^{4} \frac{\beta_{1j}}{D_{b(j)}} \frac{k_j}{u} \sin(\alpha+\theta) a_j e^{i\xi_0}\left(e^{-i\Delta_{\omega(j)}t}-1\right)\right\}. \qquad (4)$$

In the electron beam medium there are 4 branches $k_j = k_j(\omega)$ of oscillations, namely, two beams and two laser waves. Under resonant condition $\omega = \omega_+(k_0) = (\mathbf{k}_0 \mathbf{u}) - \Omega_b$ only laser wave propagating to the beam direction has maximal positive growth rate $k''$ [13]. Under the condition of appreciable amplification $k''L \geq 1$ we can omit all waves in Eq.(1) except the amplified one. The maximal value of this angle deviation is

$$\Delta\alpha_{max} = K^2 c^2 \frac{\beta_1}{D_b \gamma_0^3} \frac{k_0}{u} \sin(\alpha+\theta)\left(e^{ik''L_e}-1\right) \qquad (5)$$

For single-electron approximation (Thompson regime), Eq.(5) reduces to

$$\Delta\alpha_{max} = K^2 \left(\frac{c}{u}\right)^2 c^2 \frac{\Omega_b}{k''u} \frac{k_0}{\gamma_0^2} Xa \sin(\alpha+\theta) \frac{e^{ik''L}-1}{k''} \qquad (6)$$

Here $k''$ is given by Eq.(12) in Ref. [13]. Note that for the Thompson regime, $\Omega_b/(k''u) \ll 1$, and

$$\frac{\Omega_b}{k''u} = \frac{2^{4/3}}{\sqrt{3}} K^{-2/3} \gamma_0^{2/3} \left(\frac{k_0 u \Omega_b}{\omega^2}\right)^{1/3} \sim \omega_b^{1/3} \sim \sqrt{k''} \tag{7}$$

Hence, the angle deviation of electrons depends on the growth rate as

$$\Delta\alpha_{max} \sim \sqrt{k''} \frac{e^{ik''L} - 1}{k''} \tag{8}$$

As the growth rate diminishes $k'' \to 0$, the angle of deviation $\Delta\alpha$ goes to zero as $\Delta\alpha_{max} \sim \sqrt{k''} \to 0$.

The excess of $\Delta\alpha_{max}$ over the natural dispersion of the beam $\Delta\alpha_{beam}$ gives the threshold value of laser amplitude $a$ at the entry of the first wiggler. We rewrite formula (5) using the overall laser power $P = \frac{c}{4}(k_0 r_L)^2 |a_+|^2$ at the point of entry of the FELWI's first wiggler, namely

$$P > \frac{c}{8}\left(\frac{mc^2}{e}\right)^2 \frac{(\Delta\alpha_{beam})^2 \gamma_0^4}{2K^2 f(k''L_e)} \left(\frac{k''u}{\Omega_b}\right)^2 \tag{9}$$

The numerical value is

$$P > P_{th} = 10^9 \frac{(\Delta\alpha_{beam})^2 \gamma_0^4}{2K^2 f(k''L_e)} \left(\frac{k''u}{\Omega_b}\right)^2 \text{ (W)} \tag{10}$$

Here $f(x) = (e^x - 1)^2 / x^2$.

For calculation we consider the case of small amplification $k''L_e \sim 1$, when $f(k''L_e) \sim 1$. We assume also that $\Omega_b / (k''u) = 0.3$. Using the following values of parameters [10]: $\gamma_0 = 15$, $K = 0.635$ and $\Delta\alpha_{beam} = 5 \times 10^{-4}$ rad, we obtain value of threshold $P > P_{th} = 10^8 W$ for laser power $P$ incoming in the first undulator of FELWI. This power exceed the saturation power of the laser field for which the nonlinear regime (saturation of laser field) occurs. Therefore, the amplification regime in the first wiggler cannot be. One should decrease $\Delta\alpha_{beam}$ and/or increase $K$ (the reduction of $\gamma_0$ leads to a drop of the frequency of radiation from the optical to the radio-wave range) for lasers in the linear regime of amplification. Using the limit of laser power $\sim 10^5 \div 10^6 W/cm^2$ we obtain from formula (9) following estimation $\Delta\alpha \sim 10^{-6} rad$. This estimate coincides with the results of [13] and [14]. Note that the stable operation of FELWI demands that the value $\Delta\alpha_{max}$ should exceed the value of natural dispersion $\Delta\alpha_{beam}$ in a few times. It is doubtful that in an accelerator the natural angular dispersion of the electron beam can reach such a small value, which is significantly smaller than $10^{-6} rad$.

This means, that the regime with $k''L_e \gg 1$ or $f(k''L_e) \gg 1$ should be used of to realize the FELWI application. It can be anomalous Thompson or Raman regime amplification. For collective (Raman) regime the maximal angle of deviation is

$$\Delta\alpha_{max} = \frac{1}{2}K^2\left(\frac{c}{u}\right)^2 \frac{k_0}{\gamma_0^2} Xa\sin(\alpha+\theta)\frac{e^{ik''L}-1}{k''} \qquad (11)$$

where $X = 1 + \omega_b^2(k_0 u)/(k_0 c\Omega_b\gamma_0)$. Here $k''$ is given by Eq.(10) in Ref.[13] and, therefore, $k''$ takes the large value. The threshold power of laser is

$$P_{th} = \frac{c}{4}\left(\frac{mc^2}{e}\right)^2 \frac{(\Delta\alpha_{beam})^2 \gamma_0^4}{K^2 f(k''L_e)} \qquad (12)$$

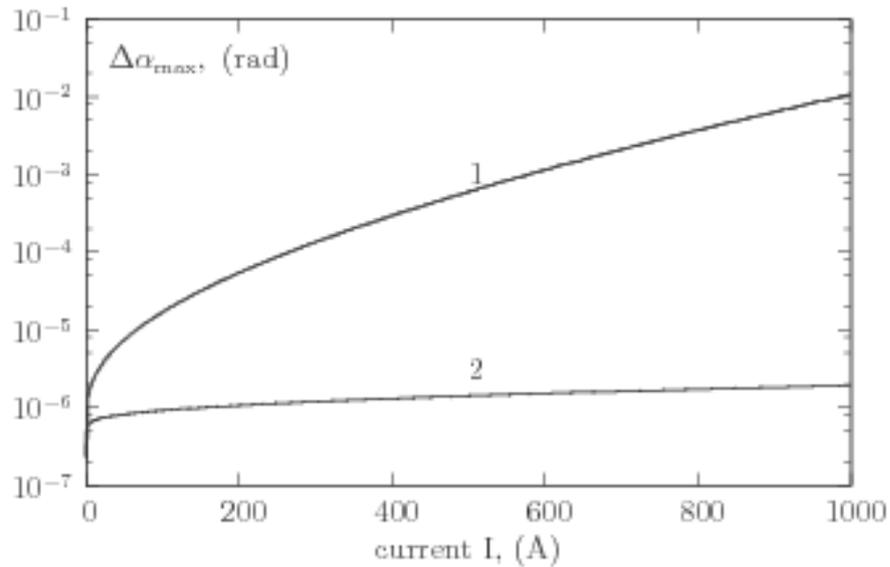

Fig.2. The deviated angle $\Delta\alpha_{max}$ as a function of a beam current I for two values of the laser beam width: line 1 corresponds r_L=1.0cm and line 2 corresponds r_L=0.1cm. Other parameters are: electron energy $\gamma = 15$, rms electron beam radius r_b=70 $\mu$m, laser wavelength

$\lambda_L = 359 \mu m$, period of the wiggler magnets $\lambda_W = 2.73 cm$, normalized wiggler field K=0.635, angle between laser and electron beam $\alpha + \beta = 0.13$.

In paper [10] the tolerance of the FELWI gain to the electron beam energy spread has been demonstrated. For this spread $\delta\gamma$ has been taken to be extremely large, namely, $\delta\gamma = 2.0$ while the emittance was $\varepsilon = 2\pi \times 10^{-6} m \cdot rad$. Simulations have been performed to obtain the dependence of the FELWI gain on the electron beam current. The results show that the gain is about 2 orders of magnitude larger than that for ordinary FEL. The simulation have been carried out with the following set of realistic electron beam and wiggler parameters that are sufficiently close to experimental situations [15 56]: electron energy $E = 29.35 MeV$ ($\gamma = 15$), emittance up to $\varepsilon = 2\pi \times 10^{-6} m \cdot rad$, rms beam radius $r_b = 70 \mu m$, laser wavelength $\lambda_L = 359 \mu m$, period of the wiggler magnets $\lambda_W = 2.73 cm$, number of magnets per section $N = 32$, normalized wiggler field $K = 0.635$, angle between laser and electron beam $\alpha + \theta = 0.13 rad$.

For our calculations we choose the same parameters and assume that the power of laser wave incoming in the first wiggler of FELWI is $P = 100W$. The results are presented in Fig. 5, which shows the angle deviation $\Delta\alpha_{max}$ as a function of a beam current $I$ for two values of the laser beam width: line 1 corresponds $r_L = 1.0 cm$ and line 2 corresponds $r_L = 0.1 cm$.

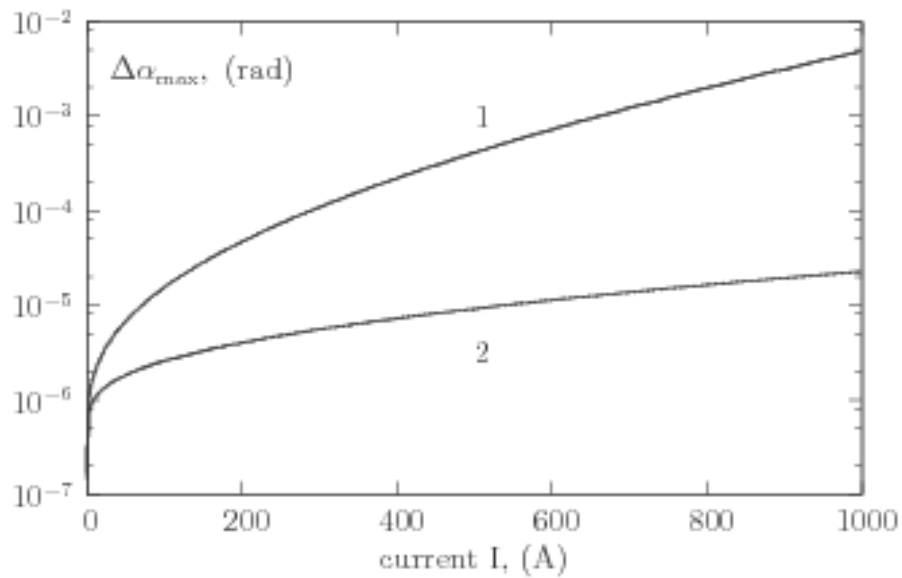

Fig.3. The deviated angle $\Delta\alpha_{max}$ as a function of a beam current I for two values of the angle between laser and electron beams: line 1 corresponds $\alpha+\beta=0.5$ rad and line 2 corresponds $\alpha+\beta=0.13$ rad. Other parameters are: electron energy $\gamma=15$, rms electron beam radius r_b=0.02cm, laser wavelength $\lambda_L=359\mu m$, period of the wiggler magnets $\lambda_W=2.73 cm$, normalized wiggler field K=0.635, r_L=1.0cm

Note that, under the condition $I>10A$, Raman amplification [12,13] takes place. The dependence of the angle deviation $\Delta\alpha_{max}$ on the laser beam width has simple explanation. On the one hand, with increasing width $r_L$ the laser amplitude drops, under the condition $P=const$. But on the other hand, the length of interaction $L_e$ increases proportional to the width $r_L$. Hence the exponential term $e^{k''L_e}$ grows.

The length of interaction can be changed with angle $\alpha+\theta$ between the electron and the laser beams. Fig.4 presents results for different values of angle $\alpha+\theta$\$: line 1 corresponds $\alpha+\theta=0.05 rad$ and line 1 corresponds $\alpha+\theta=0.13 rad$. The widths of the electron and the laser beams are $r_b=0.02cm$ and $r_L=1.0cm$, respectively. One can see that geometrical parameters, such as the widths of the electron $r_b$ and the laser $r_L$ beams, the angle between the directions of propagation of the electron and the laser beams, allow us to choose an optimal scheme for FELWI operation.

3. Conclusion

Taking into account the finite sizes of the beams, the value of the threshold laser power at the entry of the first undulator of FELWI, above which the selection of electrons via the transverse velocity in the drift region is possible, have been obtained for an FEL without inversion (FELWI). We find that an FELWI cannot operate under a weak-amplification Thompson regime, for which the spatial amplification is small: $k''L_e \ll 1$. Only a large-amplification regime, $k''L_e \gg 1$, should be used to build an FELWI. It can be either the anomalous Thompson or the Raman regime of amplification, using an electron beam with overdense current density. For an FELWI operation, the optimal angle $\alpha+\theta$ between the electron and light beams is shown to depend on the the widths of the electron $r_b$ and the laser $r_L$ beams. The mechanism of an FELWI can be realized in scheme of a ring laser.